Brewer's Conjecture and a characterization of the limits, and relationships between Consistency, Availability and Partition Tolerance in a distributed service.

Amrith Kumar,
https://hypecycles.com

Kenneth Rugg
http://bit.ly/krugg-profile

Abstract

In designing a distributed service, three desirable attributes are Consistency, Availability and Partition Tolerance. In this note we explore a framework for characterizing these three in a manner that establishes definite limits and relationships between them, and explore some implications of this characterization.

## 1. Background

At PODC 2000, Brewer[1] presented the conjecture that it is impossible to develop a web service that provides all three of the following guarantees:

- Consistency
- Availability
- Partition Tolerance.

In 2002, Nancy Lynch and Seth Gilbert presented a formal proof[2] of this conjecture and established the "CAP Theorem".

All three of these attributes are highly desirable in distributed service, and it would be ideal if one were to deliver all three guarantees. However, as demonstrated by Lynch and Gilbert, and as we shall further demonstrate here, it is not possible to develop a service that can in fact deliver all three guarantees.

Previous work, including the original conjecture, has spoken in terms of "compromising" Consistency, Availability or Partition Tolerance. We propose a mechanism by which to characterize and quantify the degree to which Consistency, Availability and Partition Tolerance of a service are achieved or compromised.

---

[1] Eric Brewer, "Towards Robust Distributed Services", PODC 2000 Keynote
[2] Nancy Lynch and Seth Gilbert, "Brewer's conjecture and the feasibility of consistent, available, partition-tolerant web services", ACM SIGACT News, Volume 33 Issue 2 (2002), pg. 51-59

## 2. Definitions

For the purposes of this note, we consider a distributed service consisting of multiple nodes, servicing requests from some repository. Client applications may submit requests to any node that is part of the distributed service.

In characterizing a service, we have focused on the attributes Consistency, Availability and Partition Tolerance. Other factors such as the amount of time taken to transmit a message from one node to another, the time taken to process a message, or the time taken waiting for some resource to become available also influence the overall performance of the service. We assume for the sake of simplicity and clarity that messages are delivered instantaneously, and that they can be processed instantaneously.

### 2.1. Consistency

The most commonly understood and expected form of consistency is called atomic consistency or linearizability[3], where it is guaranteed that there is a definite and total order on all operations, and all operations are performed as if they were performed in a single instant.

It is implied by this definition that the results of an operation will be immediately available to all subsequent operations. Another way to state this is to say that requests of the distributed service will act as if they were executing on a single node, responding to operations one at a time.

We define a service that exhibits this property as *strictly consistent*.

Ensuring this could have some undesirable costs, and in many situations a perfectly functional application may be built if the service was not strictly consistent. Such applications may be willing to tolerate some "inconsistency" in exchange for other benefits, and incorporate some mechanisms to deal with that inconsistency.

We define a service to be $T_C$ *Consistent* if every response to a request may be *inconsistent* only with respect to other requests processed by the service in the immediate preceding interval of length $T_C$.

Therefore, if a service that is $T_C$ Consistent provided a response to a request at time *T*, then all requests processed by the service up to time $T-T_C$ *would be* reflected in the response and only requests processed in the time interval $[T-T_C, T]$ *may not be* reflected in the response.

---

[3] M. Herlihy and J. Wing. Linearizability: A correctness condition for concurrent objects. ACM Transactions on Programming Languages and Systems, Volume 12, (1990) pg 463-492.

$T_C$ is therefore a measure of the extent to which the service may *compromise* consistency, or how inconsistent the results returned by the service *could possibly be*, and we formally define $T_C$ as follows:

> $T_C$ is the **maximum** amount of time that may elapse between the time when a request is processed by the service, and the time when all subsequent responses by the service will necessarily reflect the effect of that request.

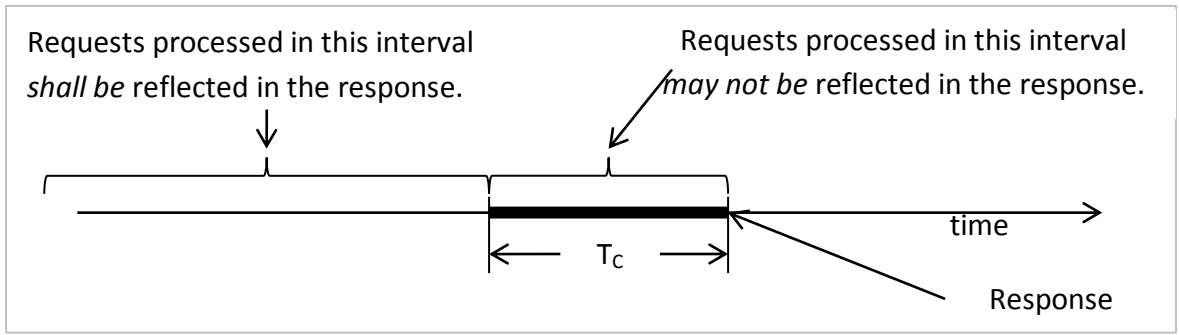

An important consequence of allowing a distributed service to return a response that does not necessarily reflect all the requests which have been processed at the time of the response is that the service may no longer be "linearized", as the effects of a request need not be propagated to all nodes in the service at the same instant.

This is illustrated in the figure below. Assume that a service that was $T_C$ consistent received three requests to write to a datum 'A' and finally a request to read the datum 'A'.

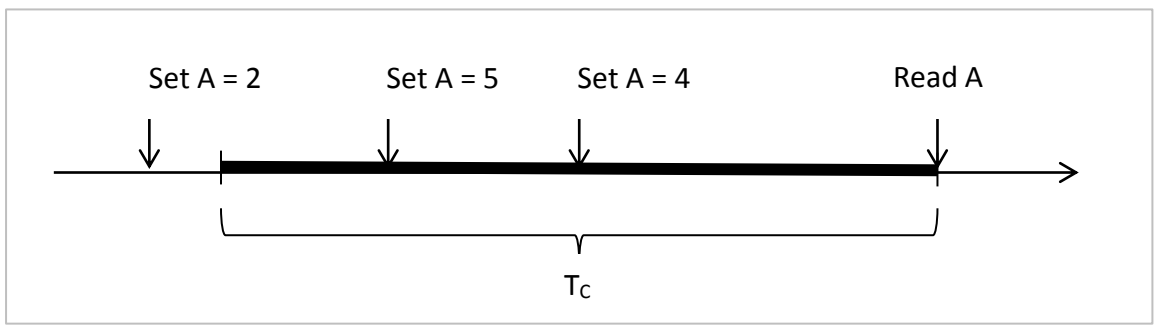

This service could return any of the values 2, 5 or 4 in response to this request. Further assuming the same scenario, it would be legitimate for the service to return values of 2, 5, 2, 2, 4, 5, 4, 4, 4, … in response to repeated requests to read datum 'A'.

We quantify the extent of data inconsistency *allowed* by a service, $T_C$ as:

- For a "Strictly Consistent" service, $T_C = 0$, and
- If $T_C > 0$, the service may generate results with some level of inconsistency. However in responding to a request, it is assumed that a response at time T would reflect at least the effect of all the requests that occurred prior to T - $T_C$. The response at time T may reflect the effects of none, some, or all of the requests that were processed after T – $T_C$.

## 2.2. Availability

A service is generally considered to be "Perfectly Available" during a period of time if all requests received by the service during that period receive an immediate response. In practice, it is not necessary that the service respond to all requests instantly, and a perfectly functional application may be built on a service that responded to requests in a timely manner.

We define the extent to which the service compromises availability, $T_A$ as:

*$T_A$ is the maximum amount of time that may elapse between the time when the service receives a request and the time when the service provides a response, again assuming that requests are processed instantaneously.*

We quantify the extent to which the service compromises availability as:

- For a "Perfectly Available" service, $T_A = 0$, and
- If $T_A > 0$, then in response to a request received at time T, the service may respond at any time before T + $T_A$.

## 2.3. Partition Tolerance

Partition Tolerance in a distributed service is the property by which the service continues to function as expected in the face of one or more (but not complete and catastrophic) discrete failures. Failures could include such things as outage on nodes, outage on network interconnects, and message loss on network interconnects, and so on.

As all software and hardware experience failures from time to time, it is possible that components of a distributed service get disconnected from each other, from time to time. In modeling a distributed service, we are primarily concerned with the inability to communicate data amongst the nodes, and the reason for this inability is not important. Consider P (t, $n_A$, $n_B$) to be a function defined as follows:

P (t, $n_A$, $n_B$) = 1 if at time 't', a node '$n_A$' in the service is able to communicate with a node '$n_B$', and

P (t, $n_A$, $n_B$) = 0 otherwise.

We could then say that, $n_A$ and $n_B$ were experiencing a partition at some time 't' if $P(t, n_A, n_B) = 0$. We depict this in the figure below.

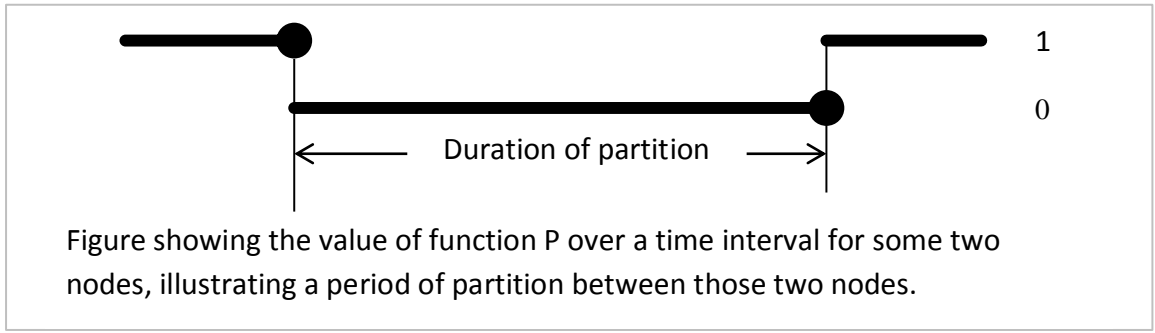

Figure showing the value of function P over a time interval for some two nodes, illustrating a period of partition between those two nodes.

We measure the amount of time for which a partition may occur between the components that comprise a service by $T_P$ where:

> $T_P$ is the maximum amount of time for which **any given node** will be unable to communicate with **some other node** in the service.

We define $T_P$ to be a measure of the extent to which the components in a service may experience a network partition, and

- If components in the service never encounter network partitions, then $T_P = 0$.
- If $T_P > 0$, then $T_P$ is the maximum interval of time for which any given node in the service is *unable* to communicate with some other node in the service.

We say that a service "that is able to perform as expected on an infrastructure that suffers network partitions" demonstrates Partition Tolerance. We can therefore quantify the Partition Tolerance of a service in terms of $T_P$ and state that:

- A service that is able to perform as expected ONLY if $T_P = 0$ is *NOT* "Partition Tolerant"
- A service that is able to perform as expected if $T_P > 0$ is "Partition Tolerant".

## 3. Relationship between $T_C$, $T_A$, and $T_P$

### 3.1. Assertion

We assert that in a service as described above,

$$T_C + T_A \geq T_P$$

In other words, a distributed service that guarantees $T_C$ Consistency, and guarantees a response in less than $T_A$, **cannot** tolerate network partitions lasting longer than $(T_C + T_A)$.

## 3.2. Proof

The proof of the assertion is provided by contradiction. Assume for the purposes of this proof that,

$$T_P > T_C + T_A$$

In other words we assume that a distributed service that guarantees results that are $T_C$ Consistent, and guarantees a response in less than $T_A$, **can** tolerate network partitions lasting longer than $(T_C + T_A)$.

We show that this is impossible.

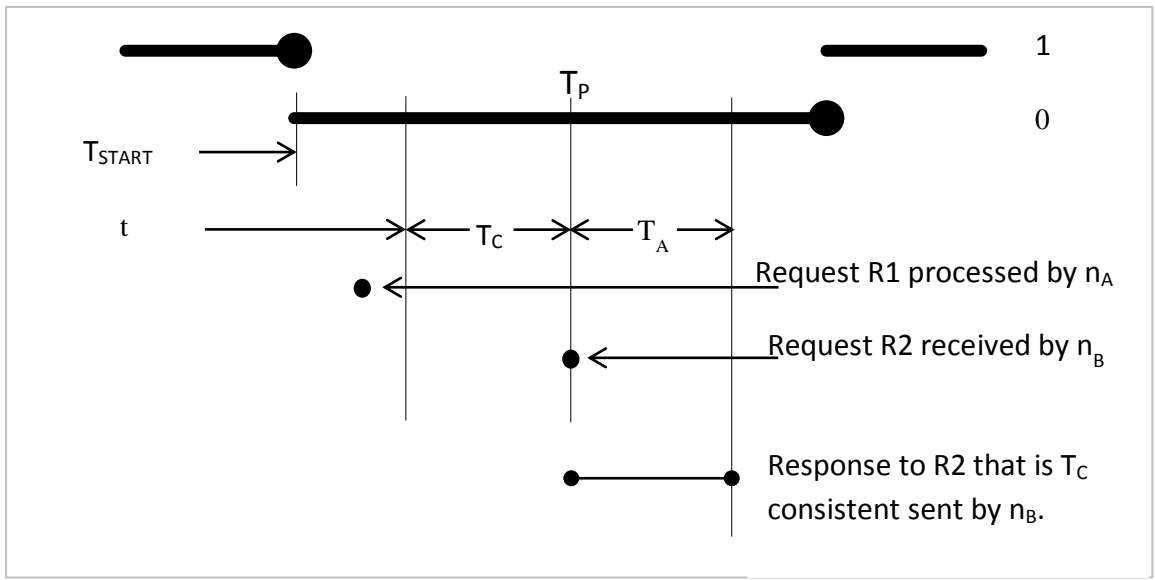

Consider the figure above depicting the period of time $T_P$ during which some two nodes ($n_A$ and $n_B$) in the service were unable to communicate with each other. As depicted in the figure, the network partition begins at $T_{START}$ and lasts for duration $T_P$.

As $T_C + T_A < T_P$, we should be able to find a time 't' such that:

- $T_{START} < t < T_{START} + T_P$, and
- $T_{START} < t + T_C + T_A < T_{START} + T_P$.

In other words, the interval of time $[t, t + T_C + T_A]$ must lie entirely within $[T_{START}, T_{START} + T_P]$, the interval when the nodes $n_A$ and $n_B$ are unable to communicate with each other.

Consider further that at some point of time in the interval $[T_{START}, t]$, a request (R1) is received by node $n_A$, and a request (R2) is received by node $n_B$ at time $t + T_C$.

As $T_C + T_A < T_P$, the node $n_B$ should be able to respond to the request R2 before the end of the network partition and that response would be $T_C$ Consistent.

In order for this to be possible, the effect of the request R1 processed by $n_A$ must have been propagated to $n_B$ before the end of the network partition between $n_A$ and $n_B$, and that is a contradiction as no messages from $n_A$ could have made it to $n_B$ before during the partition.

We therefore conclude that

$$T_C + T_A \geq T_P$$

## 4. Implications

We now consider the practical implications of the relationship that was established above.

### 4.1. Brewer's Conjecture

An immediate implication of the relationship between $T_C$, $T_A$, and $T_P$ is an alternate validation of the Brewer Conjecture.

As $T_C + T_A \geq T_P$, we can conclude the following.

- If a service is Strictly Consistent ($T_C = 0$) and Perfectly Available ($T_A = 0$), then it is *not* partition tolerant ($T_P = 0$). In other words, a Strictly Consistent and Perfectly Available service cannot also be Partition Tolerant ($T_P > 0$).
- If a service is Partition Tolerant ($T_P > 0$) then either it is not Strictly Consistent ($T_C > 0$) or the service is not Perfectly Available ($T_P > 0$). In other words, a Partition Tolerant service must compromise either Consistency or Availability.

That is Brewer's Conjecture.

### 4.2. Placing limits on Availability and Consistency in a practical service

The relationship between $T_C$, $T_A$, and $T_P$ helps establish practical limits on consistency and availability guarantees in a service.

In practice it is impossible to entirely eliminate partitions ($T_P = 0$) because failures can and will occur. While there is a high probability that the service will face short network partitions, one can design services that will have a low probability of long network partitions.

If a service is required to provide a response to each request within some amount of time $T_1$ then the best consistency guarantee that can be provided is that the service is ($T_P - T_1$) Consistent.

Similarly, if a service is required to provide a consistency guarantee of $T_2$ (i.e. that the service shall be $T_2$ Consistent), then the service cannot guarantee a response in less than ($T_P - T_2$).

Finally, if a service is required to provide a $T_C$ Consistency and a response in less than $T_A$ then the infrastructure shall guarantee that no node in the service will suffer a network partition between with any other node lasting more than $T_C + T_A$.

## 5. Conclusion

Curt Monash begins his post on "Transparent relational OLTP scale-out"[4] by saying that

> *There's a perception that, if you want (relatively) worry-free database scale-out, you need a non-relational/NoSQL strategy. That perception is false.*

As more and more people try NoSQL solutions, we are beginning to see that they are not the universal "cure-all" that they are often made out to be. As we show here in this series of blog posts, the CAP Theorem does not say that in "picking two", you have to entirely forsake the third.

As Daniel Abadi has pointed out in his PACELC blog post[5], a normally running system must make some tradeoffs between latency and consistency and what becomes interesting is how the system reacts to a network partition, does it favor availability or consistency?

---

[4] http://www.dbms2.com/2011/10/23/transparent-relational-oltp-scale-out/
[5] http://dbmsmusings.blogspot.com/2010/04/problems-with-cap-and-yahoos-little.html

For the vast majority of applications, a traditional relational database is just fine, and can provide worry-free scale-out. But, if you are really building a very large (number of nodes) database that is going to be running on an unreliable network, or is running on a widely distributed network, the implications of the CAP Theorem are more profound.

The Domain Name Service (DNS) is an example of such a system; if you think about it, it was a NoSQL database before the term was even coined. Similarly, if you are building a huge distributed search infrastructure (such as Google or Yahoo), the implications of the CAP Theorem are certainly significant. However, if you aren't operating in that rarefied atmosphere and your infrastructure is running on tens or hundreds of servers operating in a data center or the cloud, the implications of the CAP Theorem are just not that applicable to you!

To understand why this is the case, consider this. In a system with a relatively small number of nodes, or operating on a reliable network, and on reliable infrastructure, one can very readily reduce $T_p$ and thereby provide a very low floor for $T_a$ and $T_c$. However, if you are operating a very large number of nodes, or operating a highly distributed system, or one that has a high likelihood of partitioning for some other reason, you are likely to have a higher value for $T_p$, and you may in fact have to contend with that reality. Should that be the case, such as is the case with the DNS system, or the distributed search infrastructures at Google or Yahoo, you are forced into a situation where $T_a + T_c$ may in fact be pretty large.

And in situations like that, the traditional database (which is fundamentally a CA system) does face some serious challenges.

The relationship between $T_a$, $T_c$ and $T_p$ help you determine how your system will behave, and make meaningful tradeoffs between availability and consistency based on the infrastructure and environment within which the distributed database is functioning.